\documentclass[english]{IEEEtran}
\usepackage[T1]{fontenc}
\usepackage[latin9]{inputenc}
\usepackage{graphicx}
\usepackage{amssymb}
\usepackage{amsmath}
\usepackage{cite}
\usepackage{babel}

\begin{document}

\title{Obtaining the Pre-Inverse of a Power Amplifier using Iterative Learning Control}

\author{M.~Schoukens, \textit{Member}, \textit{IEEE} , J.~Hammenecker, and~A.~Cooman, \textit{Student Member}, \textit{IEEE}
\thanks{M.~Schoukens, J.~Hammenecker, and A.~Cooman are with the Department
ELEC, Vrije Universiteit Brussel (VUB), Brussels, Belgium,
e-mail: maarten.schoukens@vub.ac.be.}}

% make the title area
\maketitle

\begin{abstract}
	Telecommunication networks make extensive use of power amplifiers to broaden the coverage from transmitter to receiver. Achieving high power efficiency is challenging and comes at a price: the wanted linear performance is degraded due to nonlinear effects. To compensate for these nonlinear disturbances, existing techniques compute the pre-inverse of the power amplifier by estimation of a nonlinear model. However, the extraction of this nonlinear model is involved and requires advanced system identification techniques.
	
We used the plant inversion iterative learning control algorithm to investigate whether the nonlinear modeling step can be simplified. This paper introduces the iterative learning control framework for the pre-inverse estimation and predistortion of power amplifiers. The iterative learning control algorithm is used to obtain a high quality predistorted input for the power amplifier under study without requiring a nonlinear model of the power amplifier. In a second step a nonlinear pre-inverse model of the amplifier is obtained. Both the nonlinear and memory effects of a power amplifier can be compensated by this approach. The convergence of the iterative approach, and the predistortion results are illustrated on a simulation of a Motorola LDMOS transistor based power amplifier and a measurement example using the Chalmers RF WebLab measurement setup.
	\end{abstract}

% Note that keywords are not normally used for peerreview papers.
%\begin{IEEEkeywords}
%IEEEtran, journal, \LaTeX, paper, template.
%\end{IEEEkeywords}

\section{Introduction}
	
	To maximize their power efficiency current-day power amplifiers are most often operating close to saturation. This introduces nonlinear disturbances in the amplified signals. Such nonlinear disturbances introduce unwanted effects such as: spectral spreading, intersymbol interference, and constellation warping \cite{Karam1989,Lazzarin1994}. Digital predistortion (DPD) allows to linearize the overall system behavior by predistorting the input of the amplifier such that the desired, linearized, output is obtained.
	
	The inversion of the response of a power amplifier is of critical importance for power amplifier linearization using DPD techniques. The two most common frameworks to obtain an inverse model of an high frequency power amplifier are the direct and the indirect learning architectures \cite{Eun1997,Paaso2008}. The direct learning architecture (DLA) estimates the pre-inverse of the system directly \cite{Lim1998,Kang1998,Kang1999,Zhou2007}, often using adaptive parameter estimation routines. The indirect learning architecture (ILA) first estimates the post-inverse of the system, which is used as a pre-inverse in a second step \cite{Taijun2006}.

	This paper proposes a two-step approach instead. First, iterative learning control (ILC) \cite{Bristow2006} is used to obtain the predistorted input $u(t)$ of the power amplifier. In a second step a predistorter is estimated from the reference input $r(t)$ to the predistorted input $u(t)$ of the amplifier. This approach has the advantage that the pre-inverse estimation problem is transformed to a standard nonlinear model estimation problem, by applying the ILC algorithm first. 
	
	The ILC algorithm works on the signal level directly, it is not necessary to use the adaptive nonlinear parameter estimation approaches from the DLA framework (such as the Least Mean Square (LMS) and Recursive Least square (RLS) algorithms \cite{Widrow1976,Zhou2007}). Nor is it required to approximate the pre-inverse of the system by its post-inverse as is the case in the ILA framework \cite{Taijun2006}.
	
	\begin{figure}
		\centering
			\includegraphics[width=0.85\columnwidth]{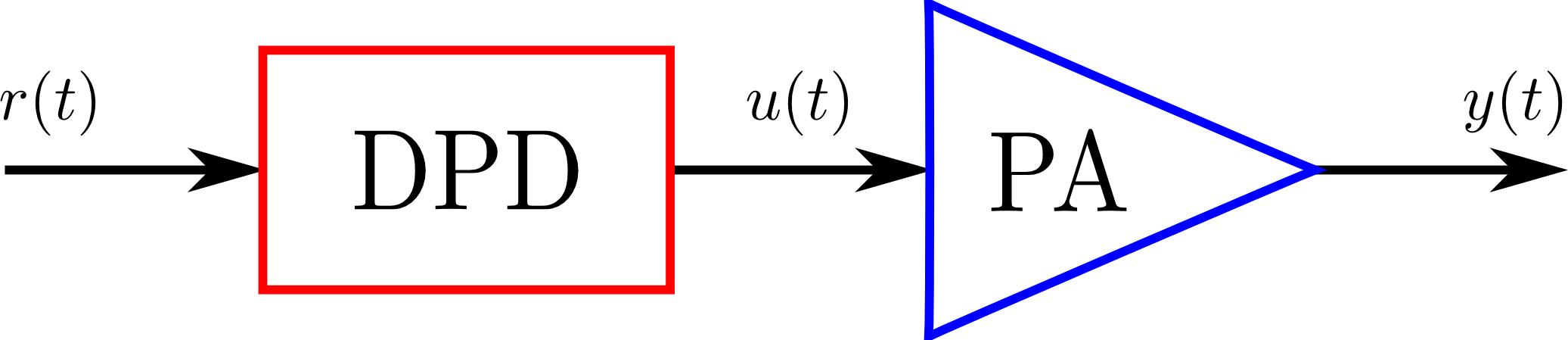}
		\caption{Simplified DPD schematic: the reference signal $r(t)$ is predistorted by the DPD, resulting in the predistorted input signal $u(t)$, such that the desired output $y(t)$ is obtained.}
		\label{fig:DPD}
	\end{figure}
	
	Such an iterative approach has been employed before to compensate for undesired memory effects that are present in a power amplifier \cite{Draxler2005}. The resulting signals can be used to act as references for adaptive control approaches, and give an indication on the importance of the memory effects in the power amplifier. Furthermore, an iterative approach can be used to obtain compensated input signals on a symbol level \cite{Deleu2014}: different blocks of input signal symbols are predistorted using an iterative algorithm. Both approaches do not compute an actual predistorter based on the iteratively predistorted signal, neither do they introduce the mature ILC framework from the control community to obtain these predistorted signals.
	
	In the remainder of this paper the DPD approach using ILC is introduced first (Section \ref{sec:DPDILC}). A high-quality simulation example in Section \ref{sec:Simulation}, and the measurement example in Section \ref{sec:Measurement} demonstrate the robustness and practical value of the proposed approach. Finally, some conclusions are drawn in Section \ref{sec:Conclusions}.

\section{DPD using ILC} \label{sec:DPDILC}
	
	\subsection{Iterative Learning Control}
	
\begin{figure}
			\centering
				\includegraphics[width=0.70\columnwidth]{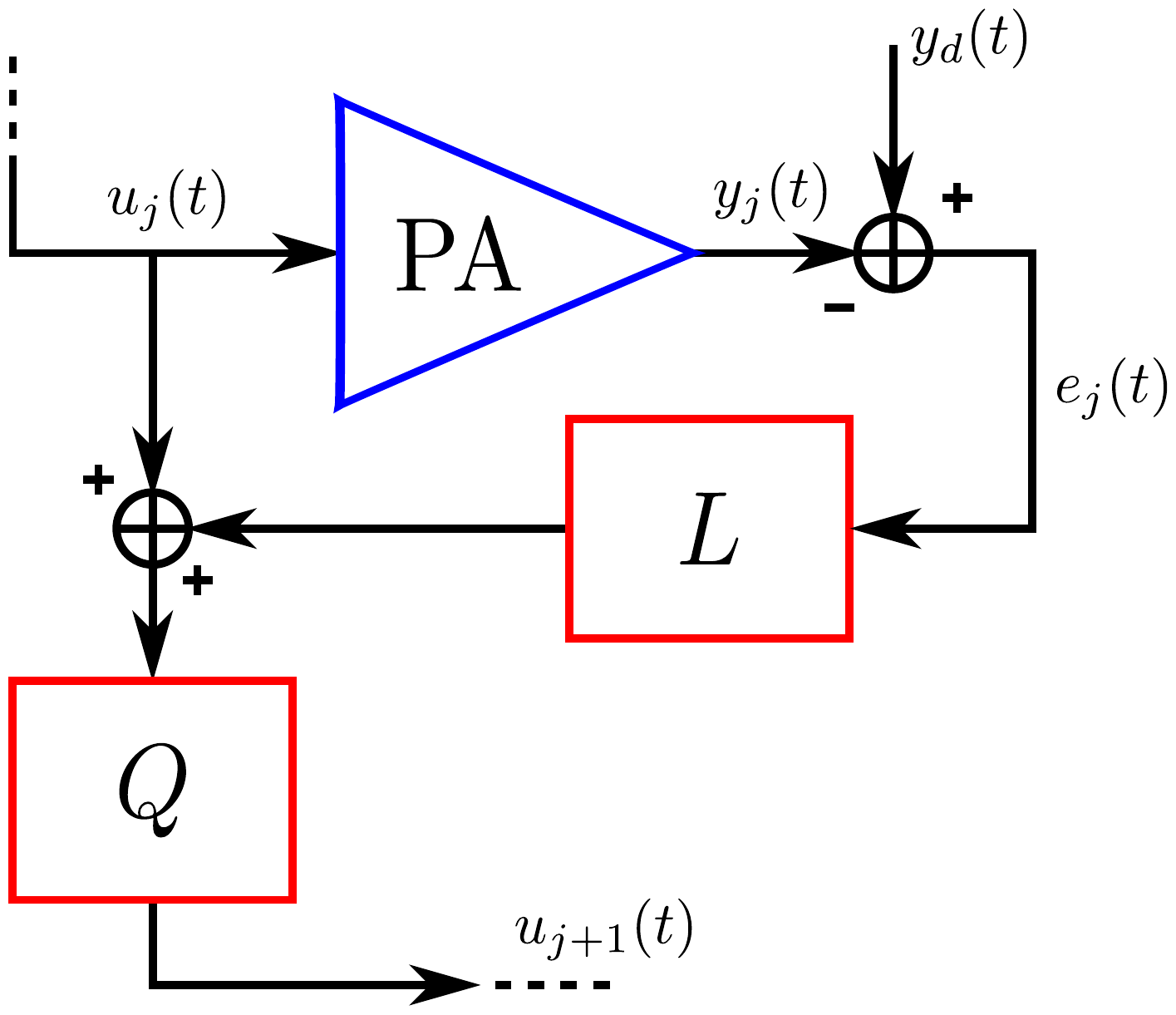}
			\caption{Iterative Learning Control schematic. The output $y_j(t)$ of iteration $j$ is subtracted from the desired output $y_d(t)$ to obtain the tracking error $e_j(t)$. This tracking error is used to obtain the input $u_{j+1}(t)$ of the next iteration using the Q-filter $Q$ and learning filter $L$.}
			\label{fig:ILC}
		\end{figure}

		Iterative learning control designs the input of a dynamic system in an iterative way to lower the tracking error of a repetitive task \cite{Bristow2006}. The ILC algorithm is visualized in Figure~\ref{fig:ILC}. The output $y_j(t)$ of iteration $j$ is subtracted from the desired output $y_d(t)$ to obtain the tracking error $e_j(t)$ of iteration $j$:
		\begin{align}
			e_j(t) = y_d(t)- y_j(t).
		\end{align}
		In the frequency domain this becomes:
		\begin{align}
			E_j(j\omega) = Y_d(j\omega)- Y_j(j\omega),
		\end{align}
		where $E_j(j\omega)$, $Y_d(j\omega)$, and $Y_j(j\omega)$ are the Fourier transform of $e_j(t)$, $y_d(t)$, and $y_j(t)$ respectively. The tracking error is used to update the input $U_{j+1}(j\omega)$ of the next iteration:
		\begin{align}
			U_{j+1}(j\omega) = Q(j\omega)\left[ U_j(j\omega) + L(j\omega) E_j(j\omega)\right],
		\end{align}
		where $Q(j\omega)$ and $L(j\omega)$ are called the Q-filter and the learning filter respectively. They determine the evolution of the tracking error over the different iterations.
		
		The goal of this paper is to obtain a good estimate of the pre-inverse of the system. Therefore, the plant inversion ILC algorithm is used \cite{Markusson2001,Markusson2001a}:
		\begin{align}
			Q(j\omega) &= 1, \\
			L(j\omega) &= \hat{P}^{-1}(j\omega),
		\end{align}
		where $\hat{P}^{-1}(j\omega)$ is an uncertain estimate of the inverse of the system.
		This results in the following ILC update equation:
		\begin{align}
			U_{j+1}(j\omega) = U_{j}(j\omega) + \hat{P}^{-1}(j\omega) E_j(j\omega).
		\end{align}
		
		Monotonous convergence over the different iterations of ILC can be obtained under weak conditions \cite{Bristow2006}, even when the inverse model $\hat{P}^{-1}(j\omega)$ of the system is only an approximation of the true system inverse.
	
	\subsection{Pre-Inverse Estimation} \label{sec:PreInverse}
		
		In a first step, a (nonparametric) linearized model of the PA is obtained. To do so, the Best Linear Approximation (BLA) of the system is estimated \cite{Pintelon2012}. This results in a linear model $G_{bla}(j\omega)$ of the nonlinear PA which is best in least squares sense, given a class of input signals (e.g. OFDM modulated input signals with a given power spectral density). The inverse of the BLA is used in the plant inversion ILC algorithm:
		\begin{align}
			\hat{P}^{-1}(j\omega) = \frac{1}{G_{bla}(j\omega)}.
		\end{align}
		
		The ILC algorithm is now used to obtain the intermediate signal labeled $u(t)$ (or equivalently $U(j\omega)$) in the simplified DPD representation in Figure \ref{fig:DPD}. The desired output is obtained as:
		\begin{align}
			Y_d(j\omega) = G(j\omega)R(j\omega).
		\end{align} 
		$G(j\omega)$ can be chosen equal to the BLA if one wants to compensate for the nonlinearities only, or it can be chosen equal to a constant gain if both the system dynamics and the system nonlinearities need to be compensated. $R(j\omega)$ is the Fourier transform of the reference signal $r(t)$, which is the input of the DPD block of Figure \ref{fig:DPD}. 
		
		The ILC algorithm is initialized with $u_0(t) = r(t)$. The algorithm typically converges in very few iterations, this is illustrated in the simulation example (see Section \ref{sec:Simulation}). For each iteration a measurement needs to be performed with the updated signal $u_{j+1}(t)$ applied as an input.
		
		The ILC algorithm results in a high-quality intermediate predistorted signal $u(t) = u_{J}(t)$, where $J$ denotes the last ILC iteration. Obtaining a pre-inverse of the PA is now reduced to estimating a high-quality model with $r(t)$ as an input and $u(t)$ as a target output. This is a standard behavioral modeling problem, which can be solved by a wide variety of approaches \cite{Ghannouchi2009}.
	
	\subsection{Strong and Weak Points of the Proposed Method} \label{sec:Advantages}
		
		The proposed approach combines some of the strong points of both the direct and the indirect learning architectures.
		
		It is shown that the exact post- and pre-inverse of a Volterra system are identical \cite{Schetzen2006}. However, behavioral models are not exact models in practice due to significant model errors and noise distortions. In this case it becomes important to make the best possible approximation. The optimal model depends on the input signals that are considered, thus if the model will be used as a pre-inverse, it should best be estimated as a pre-inverse. It should also depend as little as possible on any other approximate models of the system or its post-inverse.
		
		The direct learning architecture estimates the pre-inverse of the system directly. However, it often does so by relying on a forward model of the system \cite{Lim1998,Kang1998,Kang1999,Zhou2007}. This often leads to more involved estimation algorithms. The indirect learning architecture estimates the pre-inverse of the system based on the post-inverse of the same system \cite{Paaso2008}. It has the advantage to be more straightforward, and in many cases, it is easier to try out different model structures for the DPD.
		
		The proposed ILC estimation approach has similarities with the LMS estimation algorithm \cite{Widrow1976,Zhou2007,Draxler2005}. The main difference is that with the ILC algorithm, it is the input signal itself that is iteratively updated such that the desired output is obtained. The actual predistorter is obtained in a second step. The LMS algorithm iteratively updates the model parameters of the predistorter directly, making it more dependent on a prior choice of the structure of the pre-inverse.
		
		The proposed ILC approach provides a high-quality estimate of the predistorted input signal of the PA. The quality of that estimate is robust with respect to model errors, the quality is affected by the noise present in the measurements, and by the considered bandwidth of the signal. It allows one to estimate the pre-inverse of the system directly, without the involvement of a complex optimization scheme. It is very easy for the users to try out many different model structures for the DPD. The ILC based pre-inverse estimation will typically require more measurements than a simple post-inverse estimation algorithm due to the iterative nature of the approach. However this number of measurements still remains quite low, as is shown in Section \ref{sec:Simulation}.
		
		It can occur that the ILC algorithm does not converge to a high-quality solution when the system is very nonlinear. In this case the linear approximation is too different from the true, nonlinear, system behavior. However, this does not pose a problem for the mildly nonlinear behavior that is observed in most power amplifiers.
		
\section{Simulation Results} \label{sec:Simulation}
	\subsection{Simulation Setup}
%		The simulation example is taken from an example workspace of Keysight's Advanced Design System \cite{KeysightADS}. The amplifier is built around a Motorola LDMOS transistor similar to the MRF9742. The design is optimized for a low ACPR with an output power of $27\mathrm{dBm}$ around $850\mathrm{MHz}$ (Figure \ref{fig:PAcircuit}).
		
\begin{figure}
			\includegraphics[width=\columnwidth]{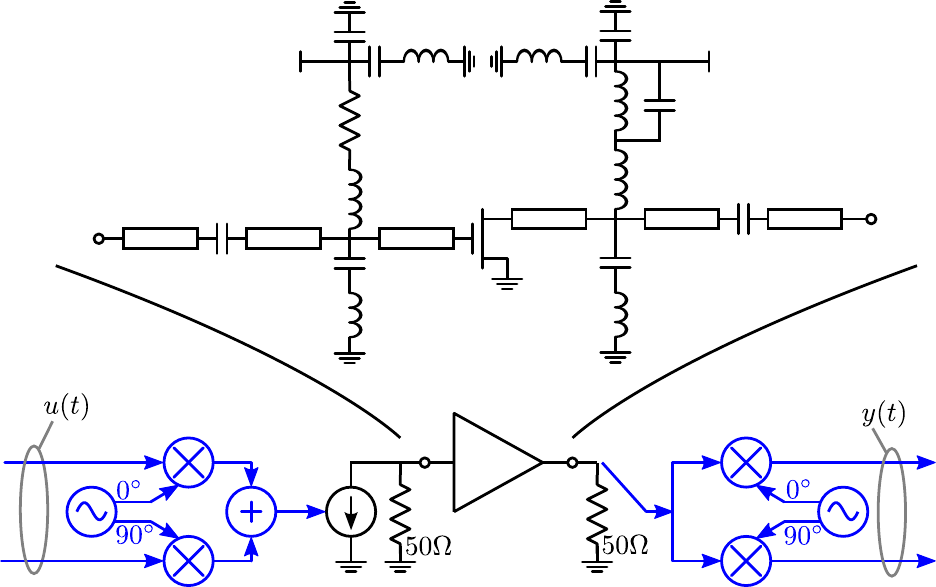}
			\caption{The PA under test is an example from Keysight's Designguide. The signals are generated and processed in a dataflow simulation. The upconverter and downconverter are both ideal blocks (shown in blue). The internals of the power amplifier are simulated using an envelope simulation.}
			\label{fig:PAcircuit}
		\end{figure}

		The simulation example is taken from an example workspace of Keysight's Advanced Design System \cite{KeysightADS}. The amplifier is built using a Motorola LDMOS transistor similar to the MRF9742 operating around $850\mathrm{MHz}$ (Figure \ref{fig:PAcircuit}).
		
		An envelope simulation combined with a dataflow simulation \cite{ADSDataFlow} was set-up to obtain the response of the circuit to modulated signals. The underlying harmonic balance simulation was performed with 10 harmonics of $850\mathrm{MHz}$. The reference signals have a bandwidth of $40$ MHz and were simulated with a fixed sample frequency of $640$ MHz. The obtained simulation results were processed in MATLAB. 
	
	\subsection{Applying ILC}
		
\begin{figure}
			\includegraphics[width=\columnwidth]{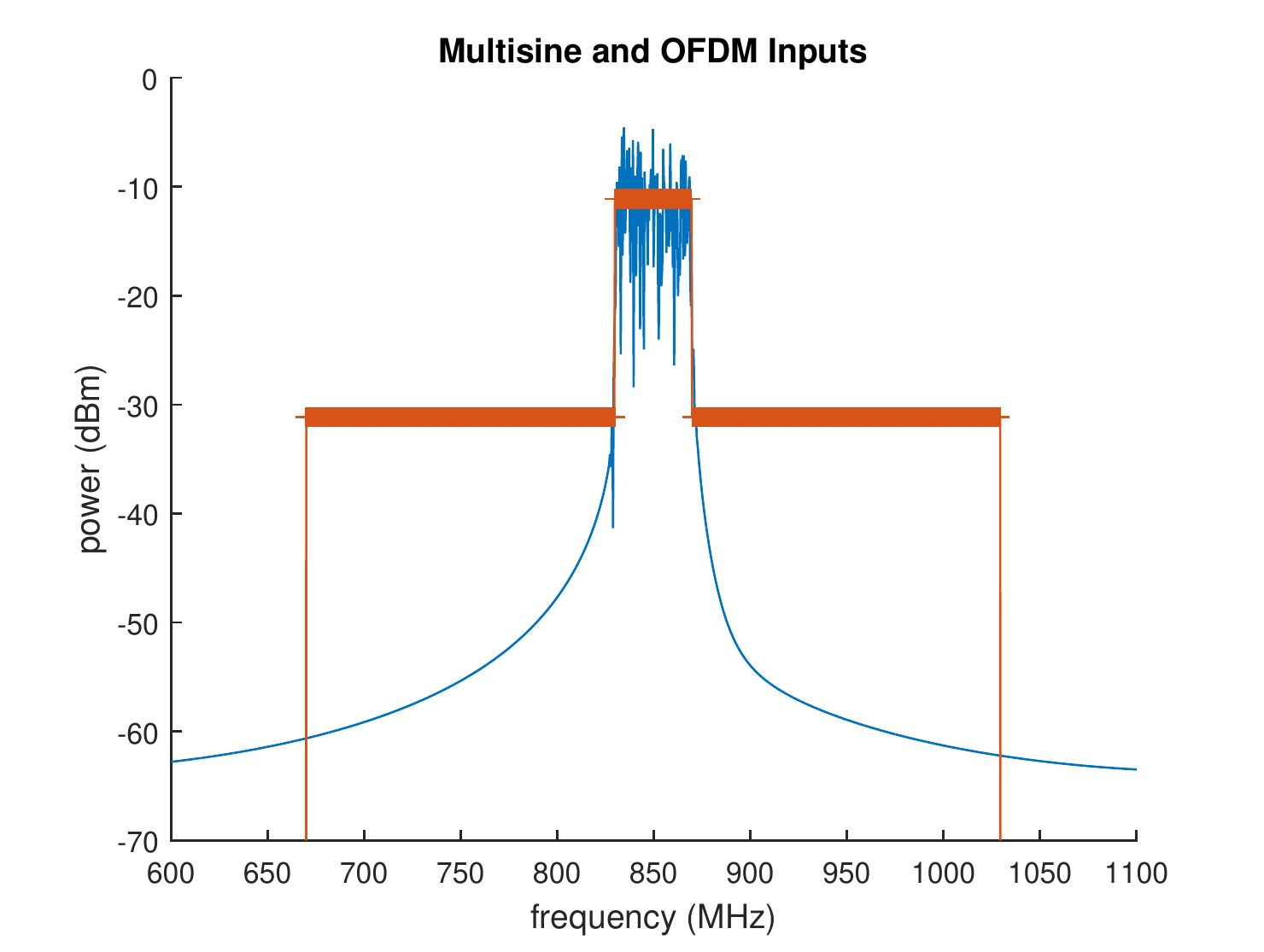}
			\caption{Multisine input signal used to estimate the BLA (red) and OFDM input signal before predistortion (blue). The multisine signal also contains energy outside the excited band of the OFDM signal such that an out-of-band linear approximation of the PA can be estimated.}
			\label{fig:Signals}
		\end{figure}

\begin{figure}
			\includegraphics[width=\columnwidth]{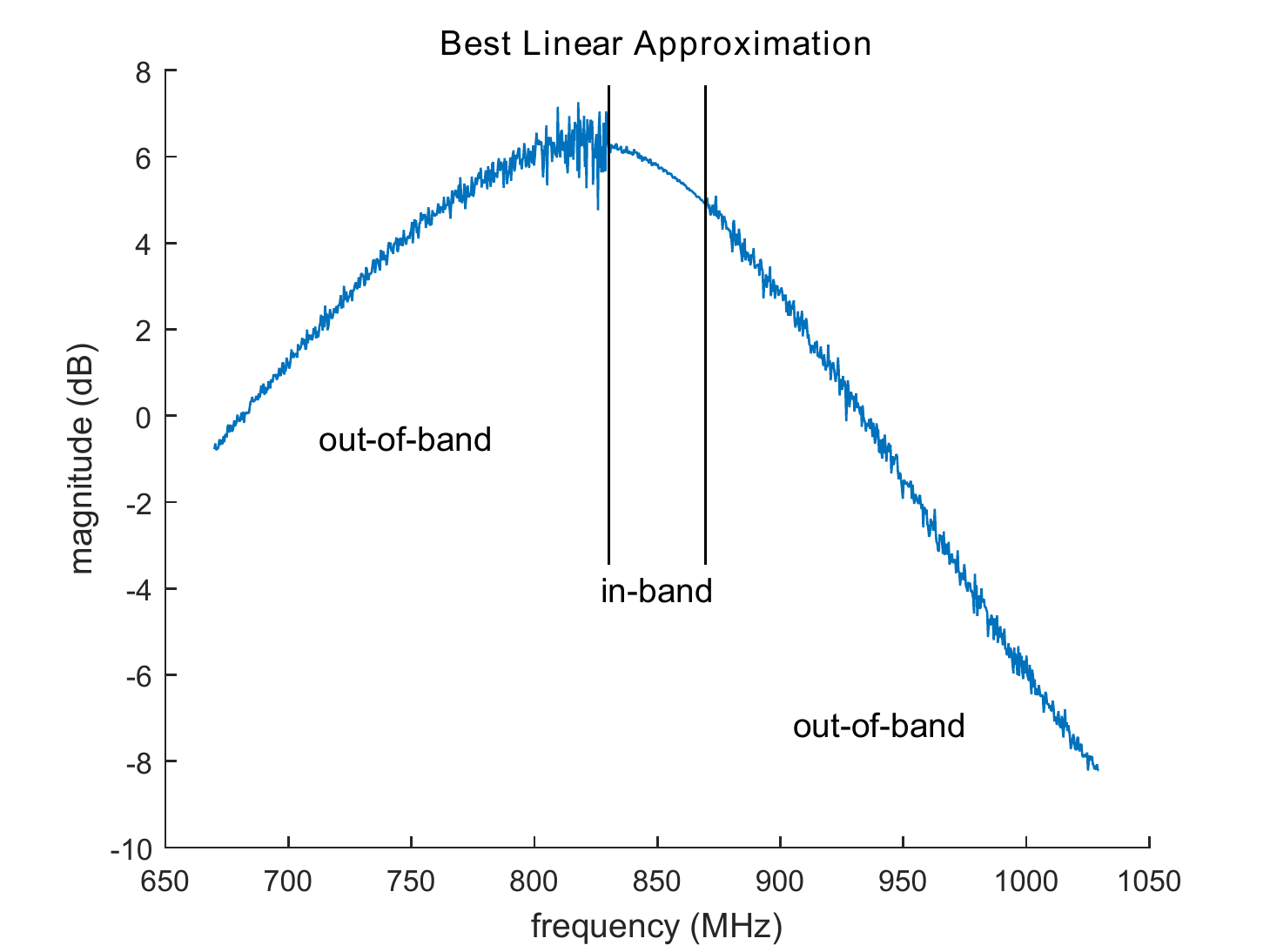}
			\caption{Estimated BLA of the power amplifier under study. Both the in-band and out-of-band FRF are estimated.}
			\label{fig:Bla}
		\end{figure}

		In a first step a linear approximation of the PA is estimated over the frequency range on which the ILC algorithm will be performed. This approximation is used as plant filter $P(j\omega)$ in the ILC. Here, the BLA is estimated over the frequency range [670, 930] MHz using multisine input signals (see Figures \ref{fig:Signals} and \ref{fig:Bla}). %Better results are obtained with the ILC algorithm for a wider frequency range. 
		
		In a second step, the ILC algorithm is applied on 20 realizations of a 1921 samples long OFDM input signal with a 40 MHz bandwidth around a 850 MHz carrier frequency. The OFDM signals have 121 excited tones, an input power of 10 dBm and a PAPR between 9.40 dB and 9.97 dB. During the ILC $G(j\omega)$ is chosen equal to the BLA, as discussed in Section \ref{sec:PreInverse}. The input power is chosen such that the nonlinear distortions are approximately 30 dBm lower than the output power. The desired output $Y_d$ is obtained as:
		\begin{align}
			Y_d(j\omega) = G_{bla}(j\omega)R(j\omega),
		\end{align}
		resulting in the compensation error:
		\begin{align}
			E_c(j\omega) = Y_d(j\omega) - Y_c(j\omega),
		\end{align}
		where $R$ is the reference input before predistortion and $Y_c$ is the predistorted output.
		
\begin{figure}
			\includegraphics[width=\columnwidth]{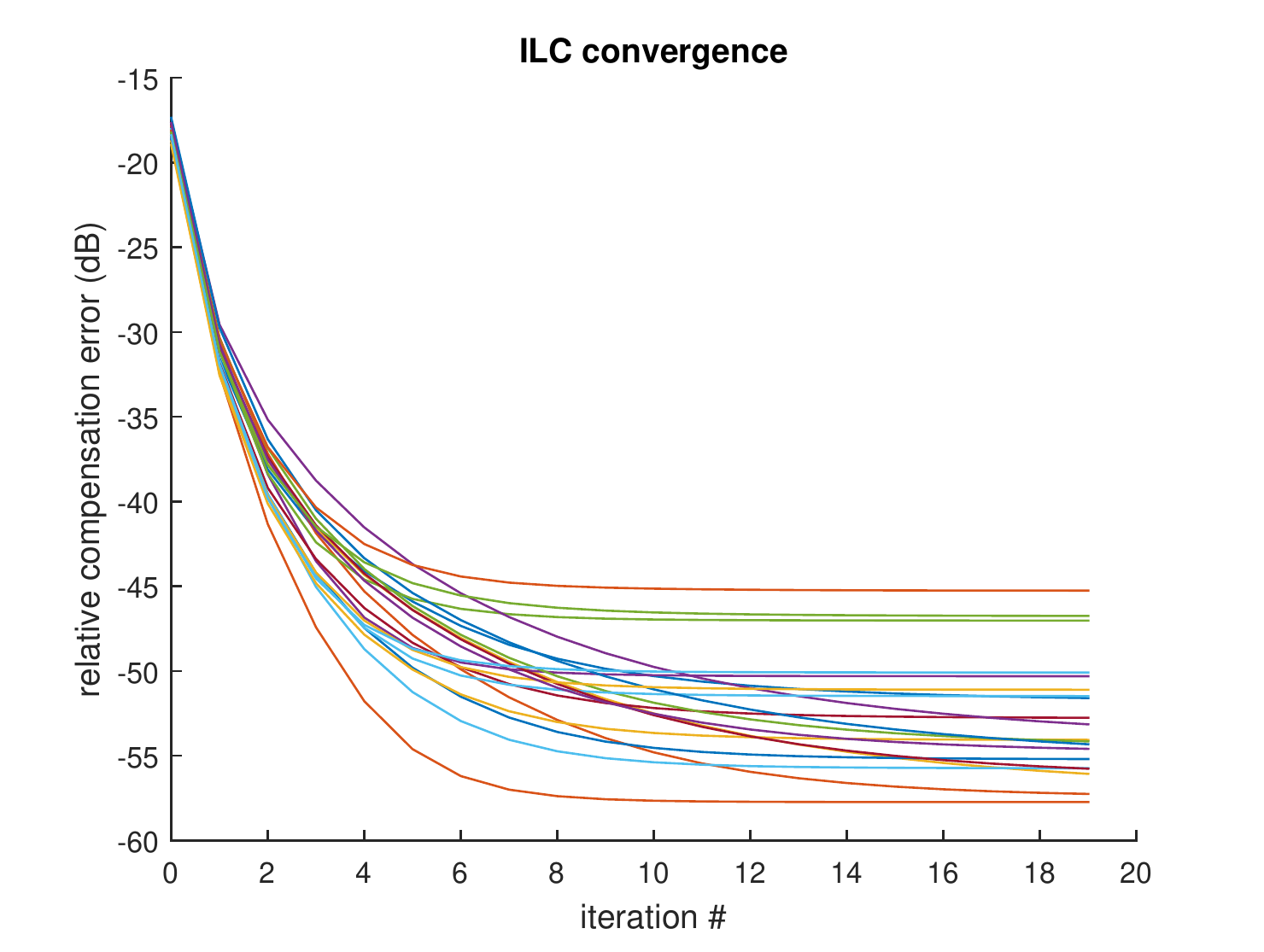}
			\caption{Convergence of the ILC algorithm for 20 different input signal realizations. Convergence is reached typically after only 6-8 iterations.}
			\label{fig:IlcConvergence}
		\end{figure}

\begin{figure}
			\includegraphics[width=\columnwidth]{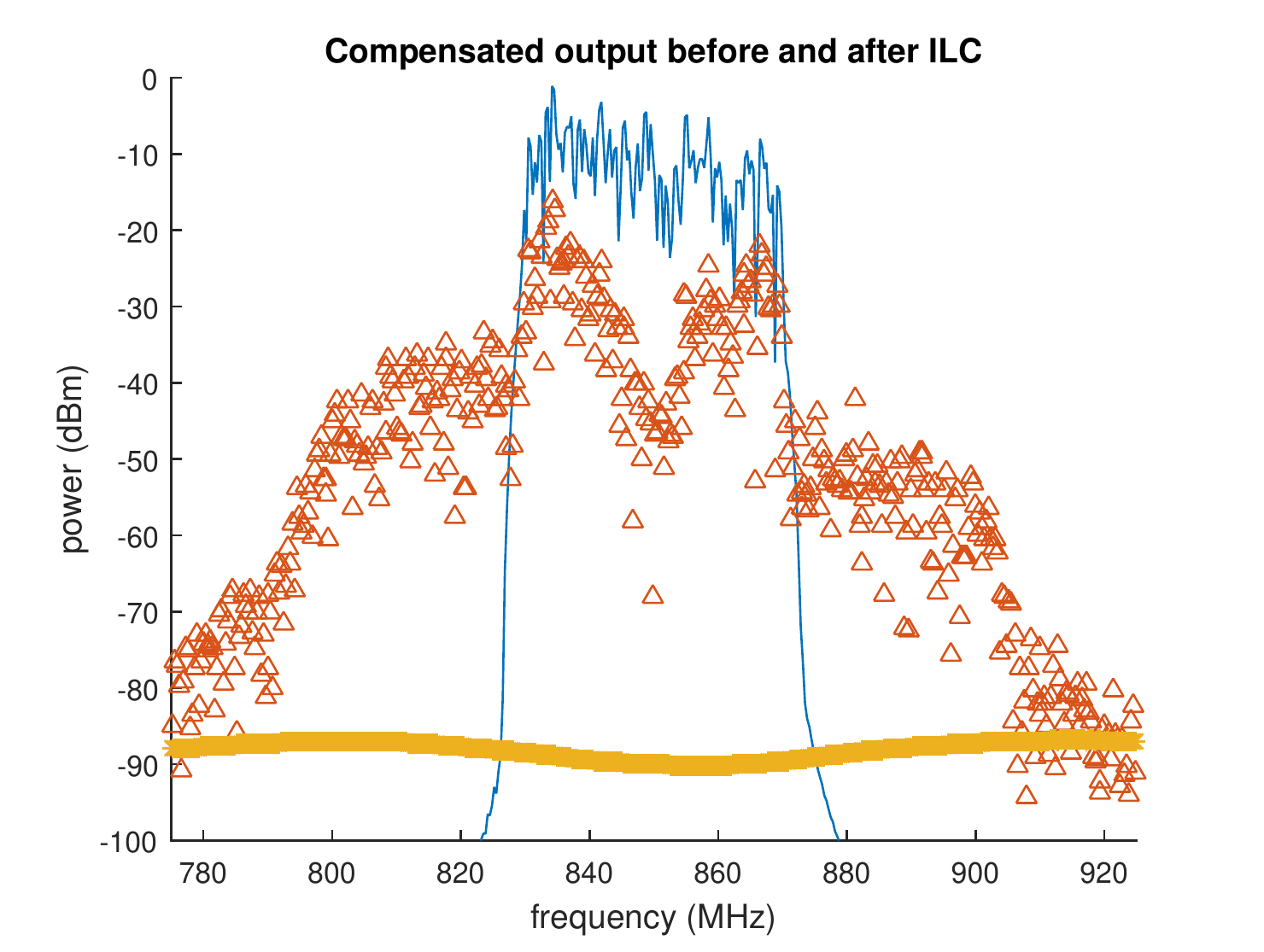}
			\caption{Frequency-domain error $E_c$ between the desired output $Y_d$ (thin full line) and the compensated output before (triangles) and after (bottom thick line) ILC.}
			\label{fig:IlcCompensated}
		\end{figure}

		The ILC obtains a high-quality estimate of the compensated input, typically after only 6-8 iterations (see Figure \ref{fig:IlcConvergence}). The compensated output, using the ILC algorithm, has an error of approximately -80 dB as can be seen in Figure \ref{fig:IlcCompensated}. The errors of the uncompensated output are both due to the nonlinear PA behavior and due to the mismatch between the desired in-band PA dynamics and the actual PA dynamics. This is an improvement of 60-70 dBm compared to the uncompensated output within the excited band, and an improvement of 50 dBm compared to the uncompensated output outside the excited frequency range. Note that this improvement is limited in real-life measurements by the noise floor of the measurement setup.
		
	\subsection{Inverse Estimation}
		
		A generalized memory polynomial (GMP) \cite{Morgan2006,Ghannouchi2009} is estimated between the reference input $r(t)$ and the predistorted input $u_{20}(t)$ (the input after 20 ILC iterations):
		\begin{align}
			u_{20}(t) &= \sum_{m=0}^{n_m} \sum_{p=0}^{n_p} \sum_{g=0}^{n_g} \alpha_{m,p,g} r(t-m) \left|r(t-m-g) \right|^p \nonumber \\
					 			&\: + \sum_{m=0}^{n_m} \sum_{p=0}^{n_p} \sum_{g=1}^{n_g} \beta_{m,p,g} r(t-m-g) \left|r(t-m) \right|^p,
		\end{align}
		 where $\alpha_{m,p,g}$ and $\beta_{m,p,g}$ are the parameters of the model, $n_m$ is the memory depth, $n_p$ is the degree of the model, and $n_g$ is the cross-term depth of the GMP.
		 
		 Both the pre-inverse and the post-inverse are estimated. The optimal model orders are determined using a cross-validation approach, 19 realizations are used for estimation and one signal realization is used as a validation sequence. This resulted in a pre-inverse GMP model with a memory depth of 5 samples of degree 7 and with a cross-term depth of 5. The post-inverse GMP model has a memory depth of 6 samples of degree 7 and with a cross-term depth of 5.
		
		Note that the presented approach is not restricted to GMP models. On the contrary, the ILC-based predistortion approach is model structure independent. The estimation of a pre-inverse boils down to a standard nonlinear model estimation problem which allows for a very flexible choice of model structures.
		
	\subsection{DPD Results}
		
		The obtained predistorters are validated using a different OFDM signal with the same properties as before, there is only one difference: the PAPR of the signal is now limited between 9.54 dB and 9.83 dB.
		
\begin{figure}
			\includegraphics[width=\columnwidth]{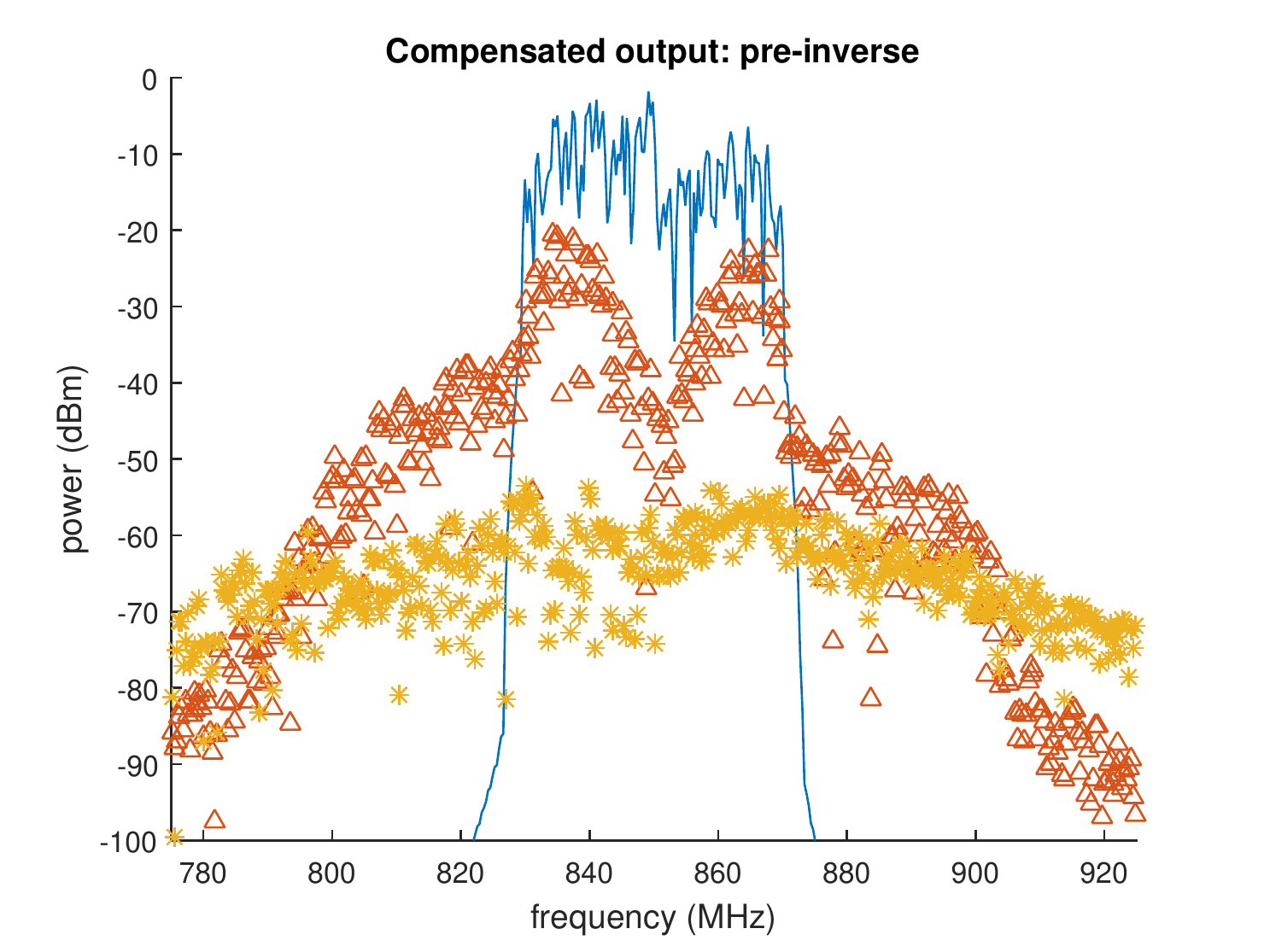}
			\caption{Frequency-domain error $E_c$ between the desired output $Y_d$ (thin full line) and the compensated output before (triangles) and after (bottom stars) predistortion.}
			\label{fig:PreInverseCompensation}
		\end{figure}

		The clear improvement of the pre-inverse ILC-based predistorter w.r.t. the uncompensated output is visualized in Figure \ref{fig:PreInverseCompensation}. Note that the obtained error in Figure \ref{fig:PreInverseCompensation} is about 20 dB higher than the error shown in Figure \ref{fig:IlcCompensated}. This illustrates that the pre-inverse ILC-based model can still be improved significantly using a more appropriate model structure. However, the search for such a model structure is not the focus of this paper.
		
		The predistortion quality is represented by the normalized rms-error between the desired output $y_{d}(t)$ and the compensated output $y_c(t)$:
		\begin{align}
			e_{rms} = \sqrt{ \frac{\sum_{t=1}^{N}\left( y_{d}(t) - y_{c}(t) \right)^2}{\sum_{t=1}^{N}\left( y_{ref}(t) \right)^2}}.
		\end{align}
		The uncompensated output results in a relative rms-error of 0.12, the post-inverse predistorter obtains a relative rms-error of 0.0089, while the pre-inverse ILC-based predistorter obtains a relative rms-error of 0.0056. 
				
\section{Measurement Example} \label{sec:Measurement}
	
\begin{figure}
		\includegraphics[width=\columnwidth]{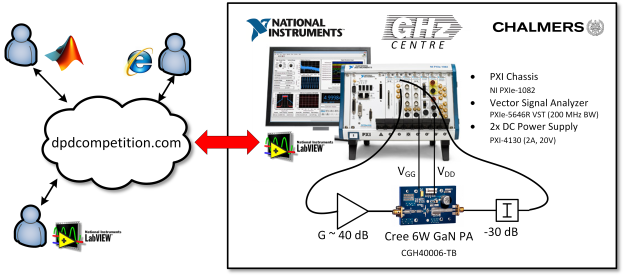}
		
		\vspace{5pt}
		
		\includegraphics[width=\columnwidth]{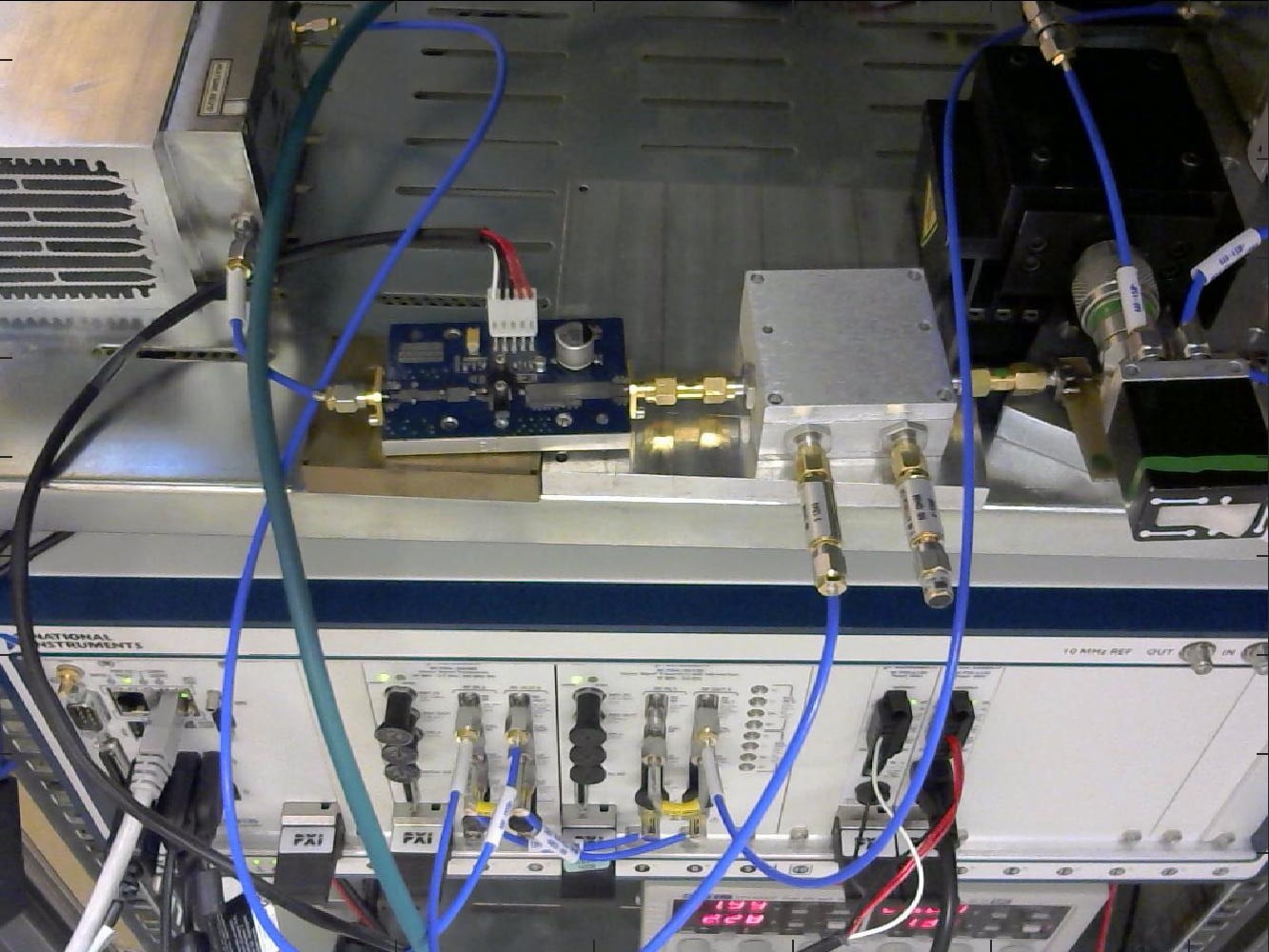}
		
		\caption{Schematic representation and picture of the RF WebLab measurement setup (source: \cite{WebLab2016}).}
		\label{fig:WebLab}
	\end{figure}

	The fast convergence of the proposed ILC-based predistortion algorithm is illustrated using the RF WebLab setup, provided by the GHz Centre, Chalmers University of Technology and National Instruments \cite{Landin2015}. A Cree CGH4006-TB \cite{CreeCGH4006P} power amplifier is studied in this section. A schematic and a picture of the measurement setup are shown in Figure \ref{fig:WebLab}. The philosophy of the RF WebLab setup is described in \cite{Landin2015}, a detailed and up to date description of the current setup is given by \cite{WebLab2016}.
	
\begin{figure}
		\includegraphics[width=\columnwidth]{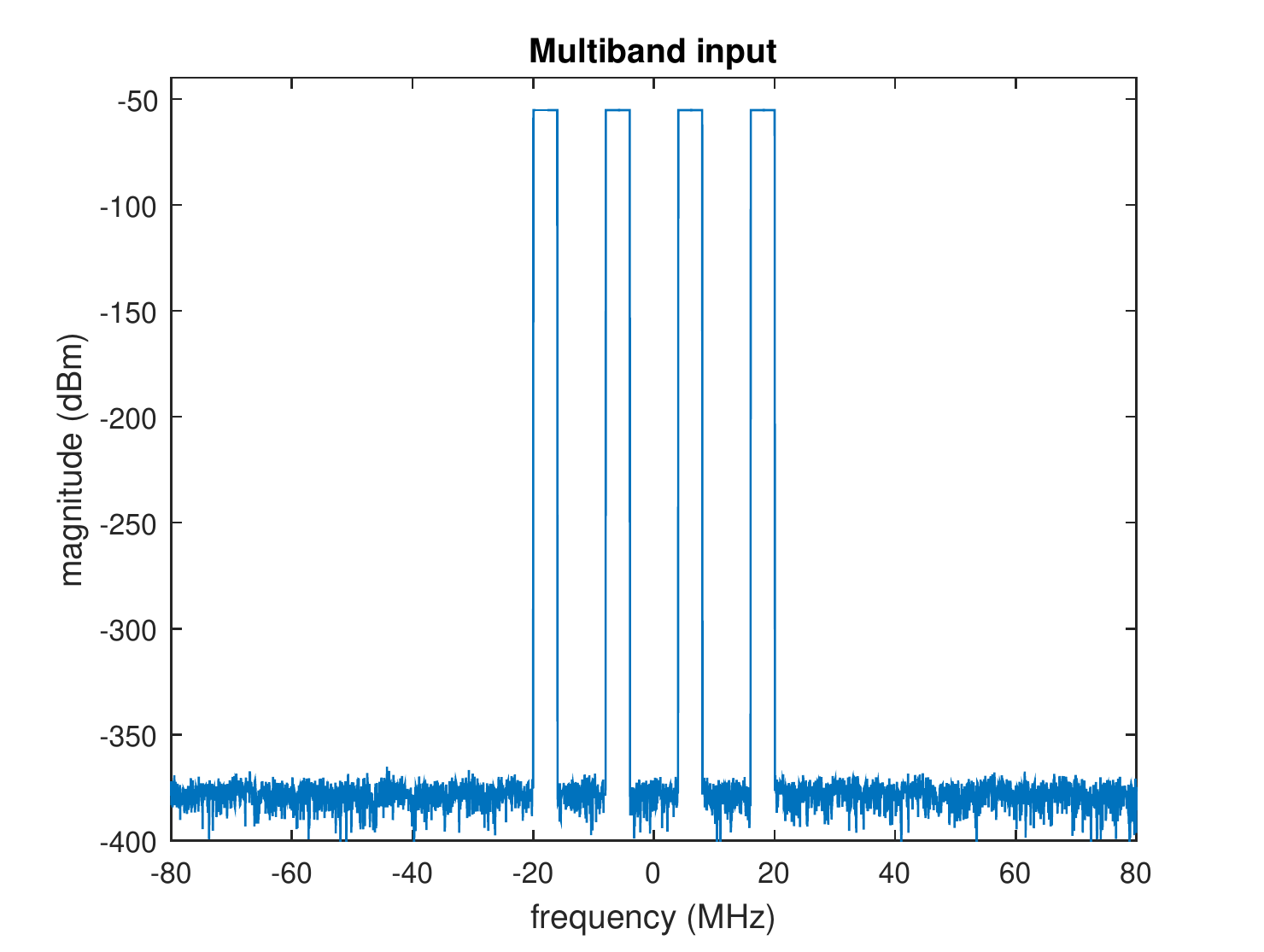}
		\caption{Multiband input signal used for the ILC-based predistortion.}
		\label{fig:SignalsWebLab}
	\end{figure}

	The power amplifier under test is excited by a multi-band input signal containing 4 excited bands of 4 MHz each, and a spacing of 8 MHz between each of them, as is shown in Figure \ref{fig:SignalsWebLab}. The input signal has an RMS input power of -30.1 dBm, a PAPR of 9.29 dBm, and a peak input power of -20.1 dBm.
	
\begin{figure}
		\includegraphics[width=\columnwidth]{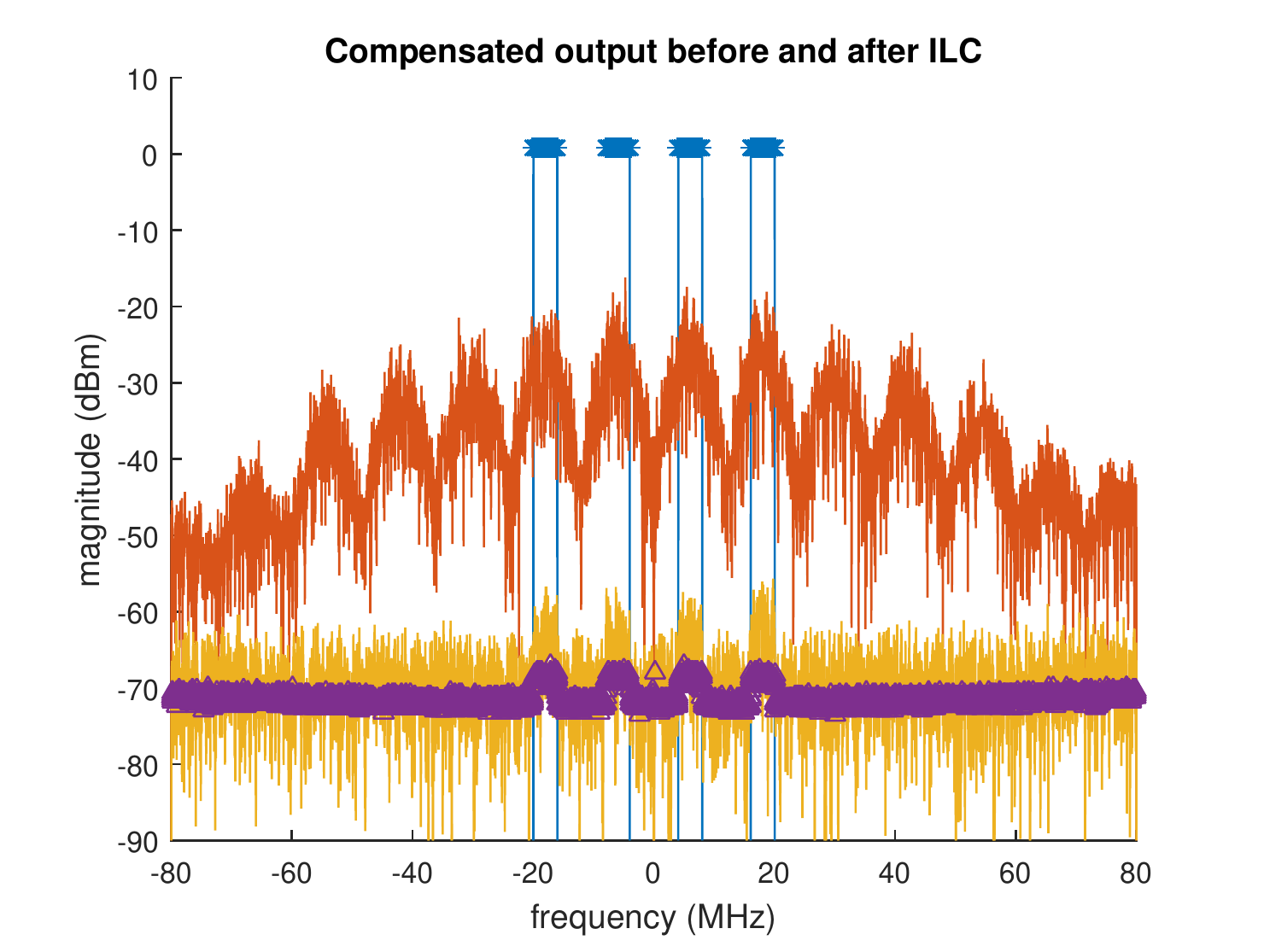}
		\caption{Frequency-domain error $E_c$ between the desired output $Y_d$ (top blue) and the compensated output before (middle red) and after (bottom orange) ILC. The estimated noise floor is depicted by the purple triangles.}
		\label{fig:IlcCompensatedWebLab}
	\end{figure}

	The ILC obtains a high-quality estimate of the compensated input, typically after only 2-3 iterations (see Figures \ref{fig:IlcCompensatedWebLab} and \ref{fig:IlcConvergence}). The compensated output, using the ILC algorithm, has an error of approximately -60 to -70 dB as can be seen in Figure \ref{fig:IlcCompensatedWebLab}. This is an improvement of 40-50 dBm compared to the uncompensated output. The compensated output is almost coinciding with the noise floor, which indicates that little further improvement can be made.
	
	\begin{figure}
		\includegraphics[width=\columnwidth]{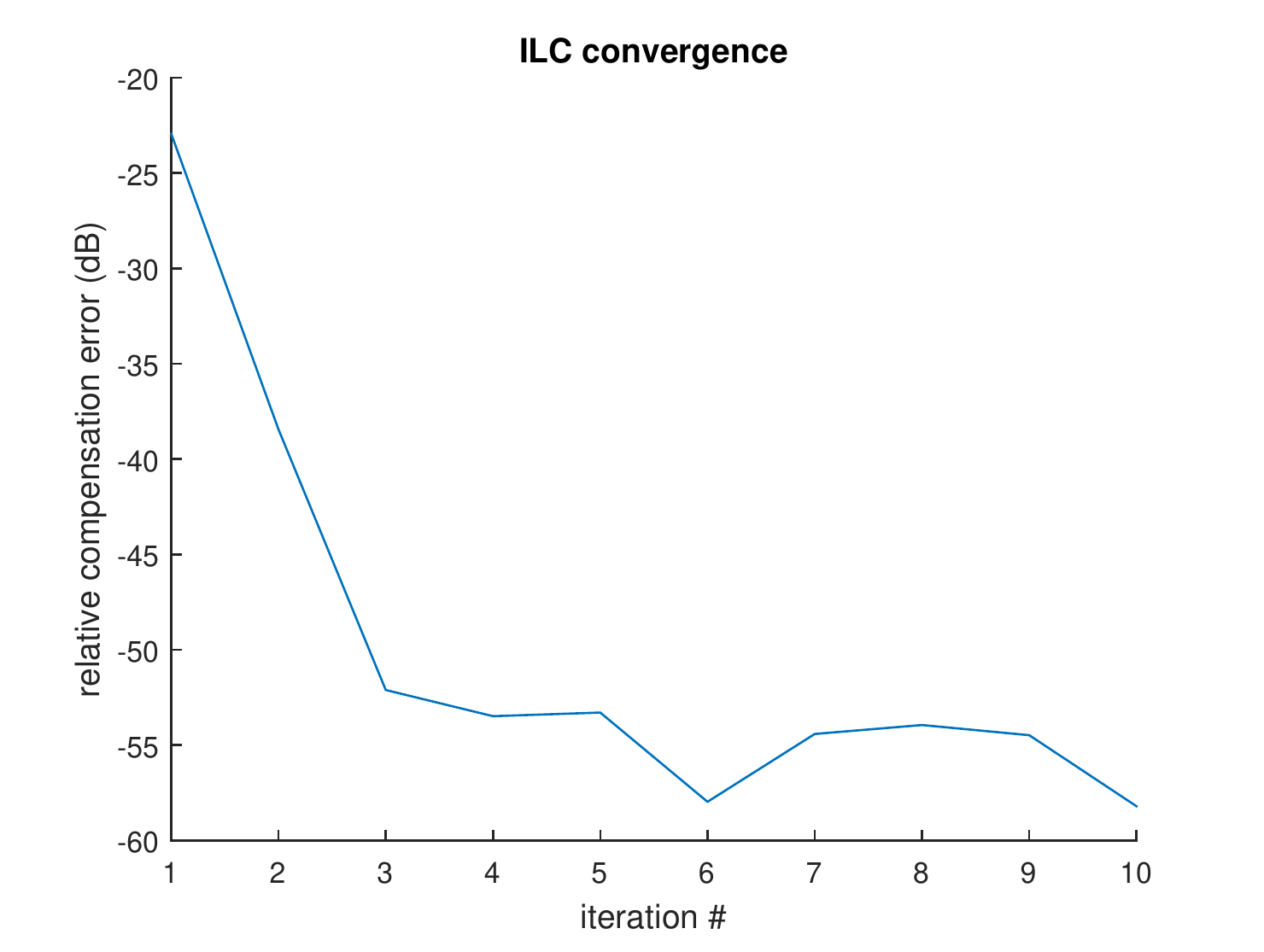}
		\caption{Convergence of the ILC algorithm. Convergence is reached after only 2-3 iterations.}
		\label{fig:IlcConvergenceWebLab}
	\end{figure}
	
\section{Conclusion} \label{sec:Conclusions}
	The proposed ILC-based predistortion algorithm allows one to obtain a high-quality predistorted input of a nonlinear power amplifier. This predistorted input can be used in a second step to estimate the pre-inverse of the the power amplifier directly, without the need for a forward model of the power amplifier. It is illustrated both with a simulation and measurement example that the ILC algorithm convergence in only a few iterations, even for a more challenging multiband input. The obtained pre-inverse based predistorter is of high quality in the simulation example, and slightly better than the post-inverse based predistorter. The ILC algorithm is independent of the chosen predistorter model structure of the pre-inverse. This allows a user to easily try out many different predistorter model structures in an easy and flexible manner.

\section*{Acknowledgments}
	This work is sponsored by the Research Foundation Flanders (FWO-Vlaanderen), the Strategic Research Program of the VUB (SRP-19), the Research Council of the VUB (OZR), the Institute for the Promotion of Innovation through Science and Technology in Flanders (IWT-Vlaanderen), and the Belgian Federal Government (IUAP VII).

% references section

\bibliographystyle{IEEEtran}
\bibliography{ReferencesLibraryV2}

% biography section
% 
% If you have an EPS/PDF photo (graphicx package needed) extra braces are
% needed around the contents of the optional argument to biography to prevent
% the LaTeX parser from getting confused when it sees the complicated
% \includegraphics command within an optional argument. (You could create
% your own custom macro containing the \includegraphics command to make things
% simpler here.)
%\begin{IEEEbiography}[{\includegraphics[width=1in,height=1.25in,clip,keepaspectratio]{mshell}}]{Michael Shell}
% or if you just want to reserve a space for a photo:

%\begin{IEEEbiography}{Michael Shell}
%Biography text here.
%\end{IEEEbiography}

% if you will not have a photo at all:
%\begin{IEEEbiographynophoto}{John Doe}
%Biography text here.
%\end{IEEEbiographynophoto}

% insert where needed to balance the two columns on the last page with
% biographies
%\newpage

%\begin{IEEEbiographynophoto}{Jane Doe}
%Biography text here.
%\end{IEEEbiographynophoto}

% You can push biographies down or up by placing
% a \vfill before or after them. The appropriate
% use of \vfill depends on what kind of text is
% on the last page and whether or not the columns
% are being equalized.

%\vfill

% Can be used to pull up biographies so that the bottom of the last one
% is flush with the other column.
%\enlargethispage{-5in}

% that's all folks
\end{document}